\documentclass[twocolumn]{aastex701}
\usepackage{amsmath}

\newcommand{\Lya}{{\rm Ly}$\alpha$ }
\newcommand{\Ha}{{\rm H}$\alpha$}
\newcommand{\Hi}{H\,{\sc i}}
\newcommand{\fesca}{$f_{\mathrm{esc}}^{\mathrm{Ly\alpha}}$}

\newcommand{\ewLya}{$\rm EW_0(Ly\alpha)$}

\defcitealias{Shimizu+26}{S26}

\begin{document}

\title{Probing Spatial Variations in IGM Transmission: Higher Ly$\alpha$ Visibility in an H$\alpha$-Selected Galaxy Overdensity at $z\simeq6.2$}

\author[orcid=0009-0001-9612-1223,gname=Shunta,sname=Shimizu]{Shunta Shimizu}
\affiliation{Department of Astronomy, School of Science, The University of Tokyo,
7-3-1 Hongo, Bunkyo, Tokyo 113-0033, Japan}
\email[show]{s.shimizu@astron.s.u-tokyo.ac.jp}

\author[orcid=0000-0003-3954-4219,gname=Nobunari,sname=Kashikawa]{Nobunari Kashikawa}
\affiliation{Department of Astronomy, School of Science, The University of Tokyo,
7-3-1 Hongo, Bunkyo, Tokyo 113-0033, Japan}
\affiliation{Research Center for the Early Universe, The University of Tokyo,
7-3-1 Hongo, Bunkyo, Tokyo 113-0033, Japan}
\email{n.kashikawa@astron.s.u-tokyo.ac.jp}

\author[orcid=0009-0007-0864-7094,gname=Junya,sname=Arita]{Junya Arita}
\affiliation{Department of Astronomy, School of Science, The University of Tokyo,
7-3-1 Hongo, Bunkyo, Tokyo 113-0033, Japan}
\email{jarita@astron.s.u-tokyo.ac.jp}

\author[orcid=0009-0000-9936-0425,gname=Ryo,sname=Emori]{Ryo Emori}
\affiliation{Department of Astronomy, School of Science, The University of Tokyo,
7-3-1 Hongo, Bunkyo, Tokyo 113-0033, Japan}
\email{emoriryo@g.ecc.u-tokyo.ac.jp}

\author[orcid=0000-0001-9840-4959,gname=Kohei,sname=Inayoshi]{Kohei Inayoshi}
\affiliation{Kavli Institute for Astronomy and Astrophysics, Peking University,
Beijing 100871, China}
\email{inayoshi.pku@gmail.com}

\author[orcid=0000-0002-7779-8677,gname='Akio K.',sname=Inoue]{Akio K. Inoue}
\affiliation{Waseda Research Institute for Science and Engineering,
Faculty of Science and Engineering, Waseda University,
3-4-1 Okubo, Shinjuku, Tokyo 169-8555, Japan}
\affiliation{Department of Physics, School of Advanced Science and Engineering,
Faculty of Science and Engineering, Waseda University,
3-4-1 Okubo, Shinjuku, Tokyo 169-8555, Japan}
\email{akinoue@aoni.waseda.jp}

\author[orcid=0000-0002-9453-0381,gname=Kei,sname=Ito]{Kei Ito}
\affiliation{Cosmic Dawn Center (DAWN), Copenhagen, Denmark}
\affiliation{DTU Space, Technical University of Denmark,
Elektrovej 327, DK2800 Kgs. Lyngby, Denmark}
\email{kei.ito.astro@gmail.com}

\author[orcid=0000-0003-3214-9128,gname=Satoshi,sname=Kikuta]{Satoshi Kikuta}
\affiliation{Department of Regional Promotion, Nara Prefectural University,
10 Funahashicho, Nara, Nara 630-8258, Japan}
\email{kikuta.astro@gmail.com}

\author[orcid=0009-0001-5183-2945,gname=Kentaro,sname=Koretomo]{Kentaro Koretomo}
\affiliation{Department of Astronomy, School of Science, The University of Tokyo,
7-3-1 Hongo, Bunkyo, Tokyo 113-0033, Japan}
\email{krkn38@g.ecc.u-tokyo.ac.jp}

\author[orcid=0000-0002-7598-5292,gname=Mariko,sname=Kubo]{Mariko Kubo}
\affiliation{Astronomical Institute, Tohoku University,
6-3 Aramaki, Aoba, Sendai, Miyagi 980-8578, Japan}
\affiliation{Department of Physics and Astronomy, School of Science,
Kwansei Gakuin University,
1 Gakuen Uegahara, Sanda, Hyogo 669-1330, Japan}
\email{markubo@kwansei.ac.jp}

\author[orcid=0000-0002-2725-302X,gname=Yongming,sname=Liang]{Yongming Liang}
\affiliation{National Astronomical Observatory of Japan,
Mitaka, Tokyo 181-8588, Japan}
\affiliation{Institute for Cosmic Ray Research, The University of Tokyo,
5-1-5 Kashiwanoha, Kashiwa, Chiba 277-8582, Japan}
\email{ymliang@icrr.u-tokyo.ac.jp}

\author[orcid=0000-0002-8857-2905,gname=Rieko,sname=Momose]{Rieko Momose}
\affiliation{National Astronomical Observatory of Japan,
Mitaka, Tokyo 181-8588, Japan}
\affiliation{Observatories of the Carnegie Science, 813 Santa Barbara Street, Pasadena, CA 91101, USA}
\affiliation{Kavli Institute for the Physics and Mathematics of the Universe (Kavli IPMU, WPI), UTIAS, The University of Tokyo, 5-1-5 Kashiwanoha, Kashiwa, Chiba 277-8583, Japan}
\email{rieko.momose@gmail.com}

\author[orcid=0000-0001-7457-8487,gname=Kentaro,sname=Nagamine]{Kentaro Nagamine}
\affiliation{Theoretical Astrophysics, Department of Earth and Space Science,
Graduate School of Science, The University of Osaka,
1-1 Machikaneyama, Toyonaka, Osaka 560-0043, Japan}
\affiliation{Theoretical Joint Research, Forefront Research Center,
Graduate School of Science, The University of Osaka,
Toyonaka, Osaka 560-0043, Japan}
\affiliation{Kavli Institute for the Physics and Mathematics of the Universe
(Kavli IPMU, WPI), UTIAS, The University of Tokyo,
5-1-5 Kashiwanoha, Kashiwa, Chiba 277-8583, Japan}
\affiliation{Department of Physics and Astronomy, University of Nevada, Las Vegas,
4505 S. Maryland Pkwy, Las Vegas, NV 89154-4002, USA}
\affiliation{Nevada Center for Astrophysics, University of Nevada, Las Vegas,
4505 S. Maryland Pkwy, Las Vegas, NV 89154-4002, USA}
\email{kn@vega.ess.sci.osaka-u.ac.jp}

\author[orcid=0000-0003-2984-6803,gname=Masafusa,sname=Onoue]{Masafusa Onoue}
\affiliation{Kavli Institute for the Physics and Mathematics of the Universe
(Kavli IPMU, WPI), UTIAS, The University of Tokyo,
5-1-5 Kashiwanoha, Kashiwa, Chiba 277-8583, Japan}
\affiliation{Waseda Institute for Advanced Study (WIAS), Waseda University,
1-21-1 Nishi-Waseda, Shinjuku, Tokyo 169-0051, Japan}
\email{masafusa.onoue@aoni.waseda.jp}

\author[orcid=0000-0003-4442-2750,gname=Rhythm,sname=Shimakawa]{Rhythm Shimakawa}
\affiliation{Waseda Institute for Advanced Study (WIAS), Waseda University,
1-21-1 Nishi-Waseda, Shinjuku, Tokyo 169-0051, Japan}
\email{rhythm.shimakawa@aoni.waseda.jp}

\author[orcid=0000-0002-0673-0632,gname=Hisakazu,sname=Uchiyama]{Hisakazu Uchiyama}
\affiliation{National Astronomical Observatory of Japan,
Mitaka, Tokyo 181-8588, Japan}
\affiliation{Department of Advanced Sciences, Faculty of Science and Engineering,
Hosei University,
3-7-2 Kajino-cho, Koganei, Tokyo 184-8584, Japan}
\email{hisakazu.uchiyama.86@hosei.ac.jp}

\begin{abstract}
The visibility of Ly$\alpha$ emission during the epoch of reionization depends on both galaxy properties and the transmission of Ly$\alpha$ photons through the intergalactic medium (IGM).
Using JWST/NIRCam F470N imaging, we identify a prominent projected overdensity of H$\alpha$ emitters (HAEs) at $z\simeq6.2$ in CEERS field.
We use Subaru/HSC NB872 imaging to measure their Ly$\alpha$ fluxes and compare the Ly$\alpha$ visibility of galaxies within $2\arcmin$ of the density center with that of galaxies outside this region.
The Ly$\alpha$-emitter (LAE) fraction is $0.41^{+0.15}_{-0.14}$ in the inner region and $0.27^{+0.19}_{-0.15}$ in the outer region.
Stacking the H$\alpha$ and Ly$\alpha$ images gives median Ly$\alpha$ escape fractions of $0.100^{+0.091}_{-0.053}$ and $0.063^{+0.130}_{-0.063}$ in the inner and outer regions, respectively.
Both measures indicate higher Ly$\alpha$ visibility in the inner region, although only at modest significance.
HAEs closer to the density center are systematically more massive and have redder UV slopes, which instead favor lower Ly$\alpha$ escape.
The environmental difference in Ly$\alpha$ visibility is therefore difficult to explain solely by variations in underlying galaxy population. 
Our results are consistent with reduced Ly$\alpha$ attenuation through a more highly ionized IGM in the overdense region, as expected if large H\,{\sc ii} bubbles preferentially form around such galaxy overdensities.
The wide-field LAE distribution provides complementary support for this scenario, as the HAE-defined density center coincides with the strongest LAE overdensity 
across the $1.4\,{\rm deg}^2$ HSC field.
These findings highlight the importance of spatial variations in IGM attenuation in shaping Ly$\alpha$ visibility near the end of reionization.
\end{abstract}



\graphicspath{{./fig_shimizu26/}}

\section{Introduction}
Cosmic reionization is the last major phase transition in the Universe, during which neutral hydrogen (\Hi) in the intergalactic medium (IGM) is ionized by ultraviolet radiation emitted by the first objects. 
Because \Lya photons resonantly scatter with \Hi, even a relatively small amount of \Hi\ can substantially alter their observed fluxes and line profiles, making \Lya emission particularly sensitive to the ionization state of the IGM (e.g., \citealt{ Miralda-Escude+98, Loeb+99, Santos+04}). 
The fraction of galaxies exhibiting \Lya emission, referred to as the Ly$\alpha$-emitter (LAE) fraction, $X_{\mathrm{LAE}}$, has therefore been used as one of several observational probes of the evolving ionization state of the IGM.
Observational studies have shown that $X_{\mathrm{LAE}}$ increases with redshift up to $z\simeq5$--$6$ and subsequently decreases at higher redshift (e.g., \citealt{Stark+11,Pentericci+11,Schenker+14,Kusakabe+20,Napolitano+24,Tang+24c,Tang+24a,Kageura+25, Jones+25,Napolitano+26a,Napolitano+26b}), consistent with increasing attenuation of \Lya emission by a progressively neutral IGM.
Similarly, the cosmic-averaged \Lya escape fraction (\fesca), which is defined as the fraction of intrinsically produced \Lya photons (or luminosity) that escape from a galaxy and are observed, is suggested to increase toward $z\simeq5$--$6$ and then decrease at $z\gtrsim6$ \citep{Hayes+11, Konno+18,Sun+23, Goovaerts+24b,Lin+24,Shimizu+26,Patrick+26}.

Reionization is expected to have been spatially inhomogeneous, with substantial spatial variations in the ionization state of the IGM \citep[e.g.,][]{Furlanetto+04,Iliev+06}. 
Spatial variations in \Lya visibility may therefore provide information complementary to its global redshift evolution.
In an inside-out reionization scenario, galaxy overdense regions, where the ionizing photon density is also high, are expected to ionize the IGM earlier and develop larger H\,{\sc ii} regions \citep[e.g.,][]{Furlanetto+04,Iliev+06}. 
Within a sufficiently large ionized region, \Lya photons can redshift sufficiently far from resonance as they propagate through the expanding Universe before encountering the surrounding \Hi\ gas, thereby reducing damping-wing absorption and increasing their transmission through the IGM \citep[e.g.,][]{Haiman+02,Santos+04,Furlanetto+06}.

Previous observations have reported frequent \Lya detections
in some galaxy overdensities at the epoch of reionization (EoR), suggesting enhanced IGM transmission in such environments (e.g., \citealt{Castellano+16,Castellano+18,Endsley+22b}). 
Recent JWST studies using a sample of galaxies selected independently of the \Lya emission have shown that the visibility of \Lya emission is high in some overdense regions (\citealt{Chen+24,Chen+26,Tang+24c,Napolitano+24}) ; however, other studies (e.g., \citealt{Leonova+25}) have reported either no clear difference compared to environments of average density, or weak \Lya emission despite the presence of significant galaxy overdensities (e.g., \citealt{Morishita+23,Li+26,Zhu+26}).
Interpreting these environmental trends is complicated because the observed \Lya emission depends not only on IGM transmission but also on environmental differences in the underlying galaxy population. 
Galaxies in overdense regions can differ systematically in physical properties that regulate the production and escape of \Lya photons through the ISM and CGM.  
Indeed, at lower redshift, \Lya visibility has also been found to decrease toward dense environments traced by \Ha\ emitters (HAEs) \citep[e.g.,][]{Shimakawa+17,Daikuhara+25}.
A galaxy overdensity does not necessarily imply enhanced \Lya visibility.
Environmental trends in \Lya emission should therefore be interpreted in the context of variations in the underlying galaxy population.

\citet{Shimizu+26} (hereafter \citetalias{Shimizu+26}) identified 84 HAEs at $z\simeq6.2$ in CEERS field \citep{Finkelstein+25} using JWST/NIRCam F470N observations and measured their \Lya emission with Subaru/HSC NB872 imaging. 
This \Ha-selected sample allows the spatial distribution of star-forming galaxies to be determined independently of \Lya emission, while the availability of both \Ha\ and \Lya measurements enables environmental comparisons using not only $X_{\rm LAE}$ but also \fesca.
In addition, the multi-wavelength data provide constraints on the physical properties of the HAE population. 

In this work, we find a prominent galaxy overdensity at $z\simeq6.2$ from the projected distribution of the \Ha-selected galaxy sample of \citetalias{Shimizu+26}.
The environmental dependence of \Lya visibility, quantified by $X_{\rm LAE}$ and $f_{\rm esc}^{\rm Ly\alpha}$, is then investigated by comparing the region around the density center with the surrounding region,
while galaxy properties are examined in relation to the projected distance from the density center.
These measurements allow us to assess whether the observed environmental trends can be explained by variations in the underlying galaxy population or instead point to spatial variations in IGM transmission.

The paper is organized as follows.
We describe the data and galaxy samples in Section~\ref{sec:data_sample} and define the projected galaxy environment from the H$\alpha$-selected population in Section~\ref{sec:densitymap}.
We then investigate the environmental variations in \Lya visibility and galaxy properties in Sections~\ref{sec:lya_visibility} and \ref{sec:galprops}, and extend the analysis to the wide-field LAE distribution around the HAE overdensity in Section~\ref{sec:wide_lae}.
In Section~\ref{sec:discussion}, we discuss the implications of these results for \Lya transmission during reionization, and in Section~\ref{sec:conclusion} we summarize our conclusions.
Throughout this paper, we assume a flat $\Lambda$CDM cosmology with $H_0=70\,\mathrm{km\,s^{-1}\,Mpc^{-1}}$, $\Omega_\mathrm{m}=0.3$, and $\Omega_\Lambda=0.7$. 
All magnitudes in this paper refer to AB magnitudes \citep{Oke}.

\section{Data and Sample}
\label{sec:data_sample}
We use the JWST/NIRCam F470N and Subaru/HSC NB872 imaging data in CEERS field employed by \citetalias{Shimizu+26}.
At $z\simeq6.2$, these filters capture \Ha\ and Ly$\alpha$, respectively (see Figure~1 of \citetalias{Shimizu+26}).
We also use the available HST and JWST multi-wavelength imaging in CEERS field (\citealt{Grogin+11,Koekemoer+11,Finkelstein+25,Wang+25}).
These data are used for the photometric-redshift selection of the HAEs and to constrain their physical properties.
The HAEs are selected based on a significant
narrow-band excess in F470N together with photometric-redshift constraints and visual inspection. 

In this study, we adopt the 84 HAEs identified by \citetalias{Shimizu+26} as the parent HAE sample.
For analyses of SED-derived galaxy properties, we use the 63 HAEs that satisfy the multi-band detection criteria of \citetalias{Shimizu+26}.
Independently, for the \Lya analysis, we use the 56 HAEs for which \citetalias{Shimizu+26} obtained reliable NB872 photometry in apertures centered on the
F470N positions, after excluding sources severely affected by contamination from nearby objects.

Throughout this work, we define LAEs as HAEs with the rest-frame \Lya equivalent width, \ewLya$\ge25$\,\AA. 
Non-LAEs are HAEs with reliable NB872 photometry that do not satisfy
this criterion, including sources with an NB872 signal-to-noise ratio
of ${\rm S/N}_{\rm NB872}\geq3$ and
${\rm EW}_0({\rm Ly}\alpha)<25$\,\AA, as well as NB872
non-detections with ${\rm S/N}_{\rm NB872}<3$.

For the environmental comparison of \Lya visibility using $X_{\rm LAE}$ and \fesca, we further restrict the NB872 sample to 21 HAEs with $-20.25<M_{\rm UV}<-18.75$, as described in Section~\ref{sec:densitymap}.
We refer the reader to \citetalias{Shimizu+26} for details of the data, sample selection, SED fitting, and NB872 photometry.

\section{ENVIRONMENT DEFINED BY H$\alpha$ emitters}
\label{sec:densitymap}

We construct a projected galaxy density map from the sky distribution of the full parent sample of 84 HAEs.
We apply an adaptive Gaussian kernel density estimator following the prescription of \citet{Abramson1982}. The pilot bandwidth is determined using edge-corrected leave-one-out likelihood cross-validation, yielding $b_0=3.28$ cMpc ($=1.36'$ at $z= 6.2$), where $b_0$ corresponds to the Gaussian $\sigma$. 
The local bandwidths are then determined from the pilot density, adopting the standard adaptive exponent of $\alpha=0.5$ (e.g., \citealt{Hatamnia+26}).
Our smoothing scale is consistent with recent high-redshift galaxy density analysis using similar adaptive KDE approaches \citep[e.g.,][]{DEugenio+25,Hatamnia+26}.

Because the boundaries of F470N footprint have a complex shape, 
we generate a random catalog that is uniformly distributed within the unmasked area to correct for the loss of effective area. 
We apply the same two-dimensional Gaussian kernel to the HAE and random catalogs to obtain the smoothed HAE surface-density field, $\Sigma_{\rm HAE}(\boldsymbol{x})$, and the random-catalog density field, $\Sigma_{\rm rand}(\boldsymbol{x})$, respectively.
Near the mask boundaries, the kernel partially extends outside the valid region, leading to an
artificial suppression of $\Sigma_{\rm HAE}$. 
The random-catalog density traces this local loss of effective area.
We therefore define the effective-area ratio as
\begin{equation}
E(\boldsymbol{x}) =
\frac{\Sigma_{\rm rand}(\boldsymbol{x})}
{\Sigma_{\rm rand,median}} .
\end{equation}
We use $E(\boldsymbol{x})$ to correct for the local area deficit and calculate the corrected surface density, $\Sigma_{\rm corr}$. 
We then obtain the overdensity parameter as
\begin{equation}
\delta(\boldsymbol{x})
=
\frac{\Sigma_{\rm corr}(\boldsymbol{x})}
{\langle\Sigma_{\rm corr}\rangle}
-1,
\end{equation}
where $\langle\Sigma_{\rm corr}\rangle$ is the mean of $\Sigma_{\rm corr}$ over the valid analysis region. 
In regions where $\Sigma_{\rm rand}$ is small, the effective-area correction amplifies noise and can produce spurious peaks. 
To mitigate this, we restrict the analysis to regions with $E(\boldsymbol{x})>0.5$.
Since CEERS covers only $\sim80\,{\rm arcmin^2}$, the resulting overdensity cannot be directly compared with measurements in other fields. 
Therefore, $\delta$ represents a relative overdensity of the $z\simeq6.2$ HAE population only within CEERS region.

\begin{figure}
    \centering
    \includegraphics[width=\columnwidth]{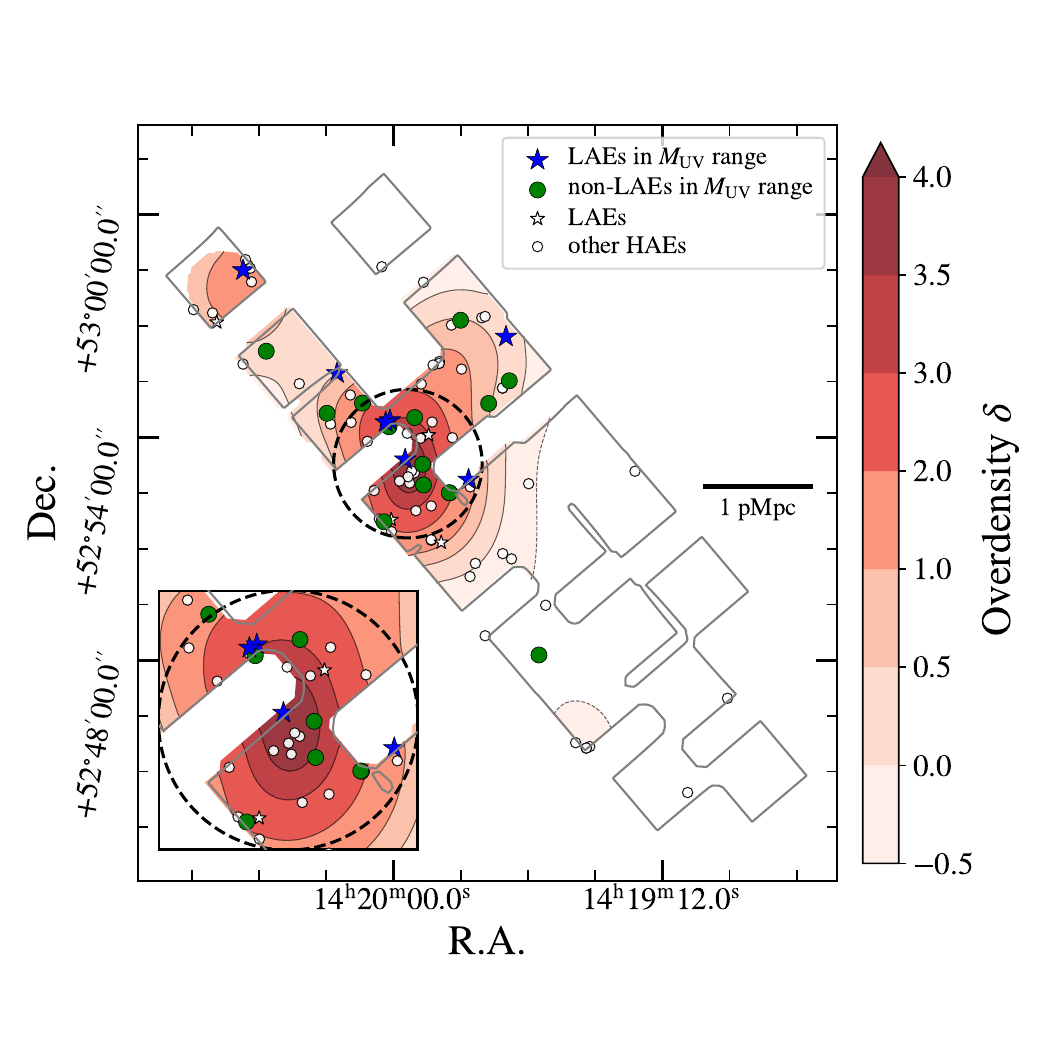}
    \caption{
    Spatial distribution of the 84 HAEs at $z\simeq6.2$ in CEERS field. The color contours show the relative overdensity $\delta$ constructed from the HAE distribution using the adaptive KDE described in Section~\ref{sec:densitymap}. Blue stars and filled green circles indicate LAEs and non-LAEs, respectively, among the 21 HAEs with $-20.25<M_{\rm UV}<-18.75$ used for the comparison of \Lya visibility, while open stars and open circles show the remaining LAEs and other HAEs, respectively. The black dashed circle is centered on the peak of the adaptive-KDE density field and has a radius of $R_{\rm cen}=2\arcmin$, corresponding to the boundary between the inner and outer regions adopted in this work. 
}
    \label{fig:HAEmap}
\end{figure}

Figure~\ref{fig:HAEmap} shows the resulting density map. 
The HAE distribution exhibits a prominent galaxy overdensity around the center of this field.
We define the position at which $\delta({\boldsymbol{x}})$ takes its maximum, $\delta=3.9$, as the center of the overdense region.
Because the local overdensity $\delta({\boldsymbol{x}})$ depends on the details of the density-map construction, including the smoothing prescription and bandwidth, we use the projected angular distance from the density center, $R_{\rm cen}$, as the environmental indicator for individual galaxies.
We confirm that the density-center position is essentially unchanged 
when the density map is constructed using only the 40 HAEs with F470N detection completeness greater than 0.8.
We note, however, that this position should be interpreted as the center of the projected overdensity of HAEs rather than as a uniquely defined physical center of the three-dimensional overdense structure.

We restrict the sample used in the following analysis to the HAEs with $-20.25 < M_{\rm UV} < -18.75$, following the conventional UV-faint luminosity interval adopted in previous measurements of $X_{\rm LAE}$ \citep[e.g.,][]{Stark+11,Pentericci+18,Kusakabe+20}.
We then apply the reliable NB872-photometry criteria described in Section~\ref{sec:data_sample}, 
resulting in 21 HAEs.
We adopt $R_{\rm cen}=2\arcmin$ as the boundary that approximately divides these galaxies into two equal-sized samples, defining 11 galaxies as the inner region and 10 galaxies as the outer region. 
A two-sample Kolmogorov--Smirnov (KS) test shows no statistically
significant difference between their $M_{\rm UV}$ distributions.
We confirm that the results are insensitive to the precise choice of this boundary: the sample membership is unchanged for $1\farcm7\leq R_{\rm cen,cut}\leq2\farcm0$, and the qualitative trends in both $X_{\rm LAE}$ and the median \fesca\ remain unchanged when the boundary is extended to $2\farcm5$.

\section{\boldmath Ly$\alpha$ visibility of H$\alpha$ emitters}
\label{sec:lya_visibility}
\subsection{LAE fraction}
\label{sec:lae_fraction}

We measure $X_{\mathrm{LAE}}$ inside and outside the overdense region using 21 HAEs defined in Section~\ref{sec:densitymap}.
Following \citetalias{Shimizu+26}, the continuum-subtracted \Lya flux is derived from the NB872 flux using the UV continuum flux and UV slope $\beta$ to
estimate the continuum contribution at 1216\,\AA.
\ewLya\ is then calculated from the \Lya flux and the continuum flux density at 1216\,\AA.
The uncertainty in $X_{\rm LAE}$ is estimated using 50,000 Monte Carlo realizations.
In each realization, the NB872 flux, UV continuum flux, and $\beta$ are perturbed according to their observational uncertainties, and 
\ewLya\ are recalculated.
HAEs satisfying \ewLya\ $\ge25$\,\AA\ in each realization are then classified as LAEs.
Because the number of galaxies in each environmental subsample is small, we additionally account for the uncertainty associated with the binomial distribution from the finite sample size. 
Following the Bayesian treatment of small-sample binomial fractions adopted in \citet{Cameron+11}, for each Monte Carlo realization, we draw $X_{\rm LAE}$ from the Jeffreys-prior Beta posterior,
\begin{equation}
X_{{\rm LAE},i}
\sim
{\rm Beta}
\left(
K_i+\frac{1}{2},
N-K_i+\frac{1}{2}
\right),
\end{equation}
where $K_i$ is the number of galaxies satisfying the \ewLya\ threshold in realization $i$, and $N$ is the total number of HAEs in the corresponding environmental sample. 
The posterior draws over all realizations are used to estimate the uncertainty in $X_{\rm LAE}$, including both the \Lya measurement uncertainties and the binomial uncertainty.

\begin{figure}
    \centering
    \includegraphics[width=\columnwidth]{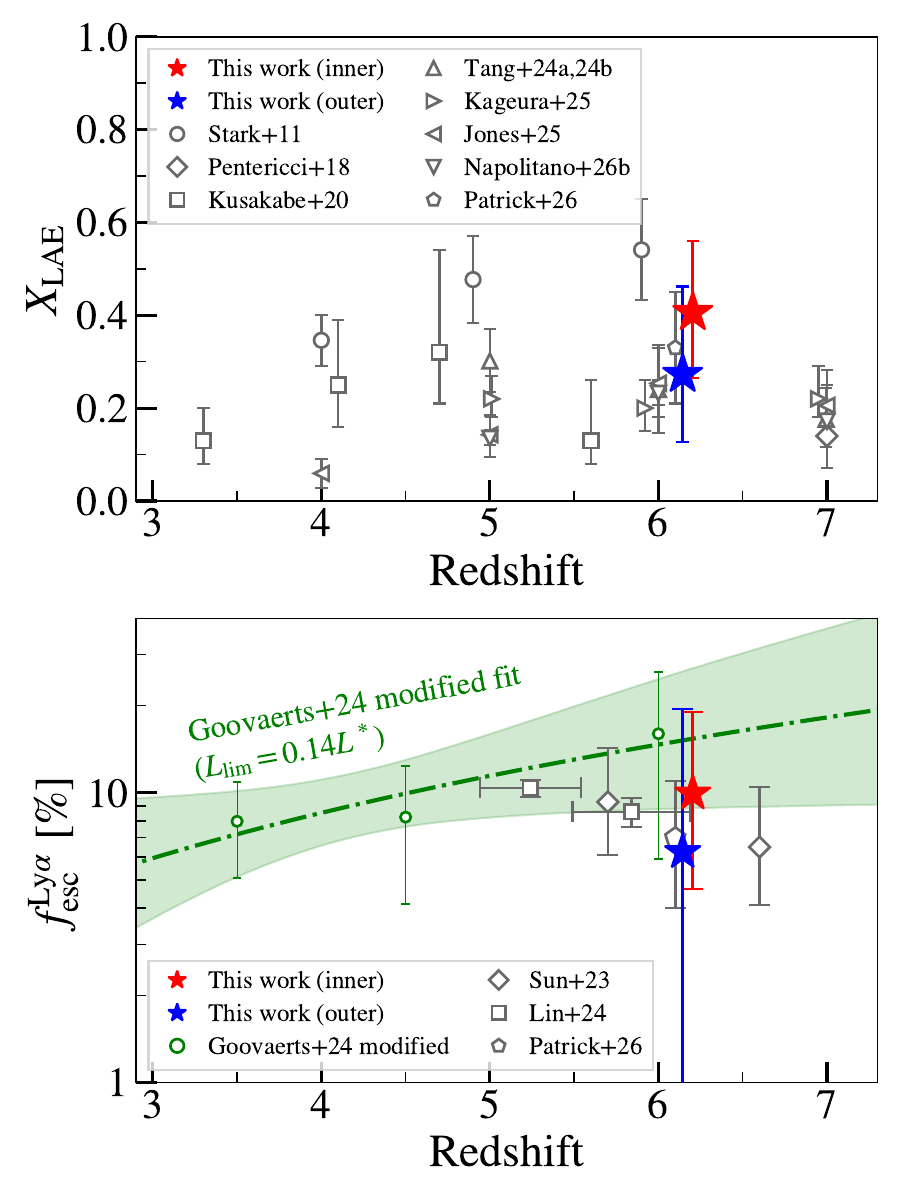}
    \caption{Redshift evolution of \Lya visibility for HAEs with $-20.25<M_{\rm UV}<-18.75$. Red and blue stars indicate our measurements in the inner and outer regions, respectively, and are slightly offset horizontally for clarity. The upper panel shows the redshift evolution of $X_{\rm LAE}$
    and gray symbols show previous measurements from \citet{Stark+11}, \citet{Pentericci+18}, \citet{Kusakabe+20}, \citet{Tang+24c,Tang+24a}, \citet{Kageura+25}, \citet{Jones+25}, \citet{Napolitano+26b}, and \citet{Patrick+26}. The lower panel shows the redshift evolution of \fesca, together with previous measurements from \citet{Sun+23}, \citet{Lin+24}, and \citet{Patrick+26} shown by gray symbols. 
    The green circles show the \fesca\ values recalculated following \citet{Goovaerts+24b} by integrating the UV and \Lya luminosity functions to a common limit of $L_{\rm lim}=0.14L^*$. The green dash-dotted curve and shaded region show the power-law fit to these values and its 16th--84th percentile uncertainty, respectively.
    }
    \label{fig:lya_evolution}
\end{figure}

As a result, we obtain $X_{\rm LAE}=0.41^{+0.15}_{-0.14}$ in the inner region and $0.27^{+0.19}_{-0.15}$ in the outer region.
Thus, $X_{\rm LAE}$ is higher inside the HAE overdensity, although the statistical significance of the difference remains moderate according to Fisher's exact test ($p\sim0.38$). 
The upper panel of Figure~\ref{fig:lya_evolution} shows the redshift evolution of $X_{\rm LAE}$ for galaxies selected with the same criteria, $-20.25<M_{\rm UV}<-18.75$ and ${\rm EW_0(Ly\alpha)}\ge25$ \AA\ \citep{Stark+11,Pentericci+18,Kusakabe+20,Tang+24c,Tang+24a,Kageura+25,Jones+25,Napolitano+26b,Patrick+26}. 
The measurement in the outer region is consistent with recent measurements at $z\simeq6$, whereas the measurement in the inner region is higher,
although their uncertainties substantially overlap.

\subsection{Ly$\alpha$ escape fraction}
\label{sec:fesc_environment}

In addition to $X_{\rm LAE}$, we derive the median \fesca\ for the inner and outer regions via stacking.
Previous studies have reported a dependence of \fesca\ on $M_{\rm UV}$, with higher \fesca\ towards fainter UV magnitudes (e.g., \citealt{Tang+24a,Goovaerts+24b}). 
We therefore restrict the stacking analysis to the same range of $-20.25<M_{\rm UV}<-18.75$ adopted for $X_{\rm LAE}$, reducing the effect of differences in the intrinsic UV-luminosity distribution when comparing environmental variations in \Lya visibility.

We use the same stacking procedure as \citetalias{Shimizu+26}, including the completeness weights based on the F470N magnitudes. 
We construct the median stacked F470N, NB872, and continuum images, independently for the inner and outer regions, and derive the \Ha\ and \Lya luminosities. 
The photometry, continuum subtraction, dust correction of \Ha, correction for the differential filter transmission between NB872 and F470N, and estimation of the stacking uncertainty follow \citetalias{Shimizu+26}.
The median \fesca\ is calculated from the observed \Lya luminosity, $L_{\rm Ly\alpha}^{\rm obs}$ and the dust-corrected intrinsic \Ha\ luminosity, $L_{\rm H\alpha}^{\rm int}$, assuming a typical gas condition (a temperature of $T_{\rm e}\sim10^4\,{\rm K}$ and an electron density of $n_{\rm e}\sim350\,{\rm cm^{-3}}$) and Case B recombination \citep{Hummer},
\begin{equation}
f_{\rm esc}^{\rm Ly\alpha}
=
\frac{L_{\rm Ly\alpha}^{\rm obs}}
{8.7\,L_{\rm H\alpha}^{\rm int}}.
\end{equation}

Stacking the 11 HAEs in the inner region yields $f_{\rm esc}^{\rm Ly\alpha}=0.100^{+0.091}_{-0.053}$, while the stack of the 10 HAEs in the outer region gives $f_{\rm esc}^{\rm Ly\alpha}=0.063^{+0.13}_{-0.063}$. 
The median \fesca\ follows the same trend as $X_{\rm LAE}$, with a higher value in the inner region. 
However, the difference is only at modest significance due to the large uncertainty in the outer-region measurement.
Although neither measurement alone provides a significant detection of environmental dependence, both $X_{\rm LAE}$ and the stacked \fesca\ consistently show higher \Lya visibility in the inner region. 

The lower panel of Figure~\ref{fig:lya_evolution} compares our \fesca\ measurements with those in previous studies \citep{Sun+23,Lin+24,Goovaerts+24b,Patrick+26}. 
For a luminosity-matched comparison with the global \fesca\ evolution, we follow the approach of \citet{Goovaerts+24b}. 
We integrate the UV luminosity function \citep{Bouwens+22} and modified Schechter \Lya luminosity functions of \citet{Thai+23} down to a common limit of $L_{\rm lim}=0.14L^\ast$, corresponding approximately to the faint-end limit of our stacking sample, $M_{\rm UV}=-18.75$ \citep{Bouwens+22}. 
We then fit the resulting three measurements at $z=3.5$, 4.5, and 6.0 with a simple power law, $f_{\rm esc}^{\rm Ly\alpha}=A(1+z)^\xi$, obtaining $A=6.35\times10^{-3}$ and $\xi=1.61$.
The uncertainty in the fitted evolution is estimated using Monte Carlo simulations, with the 16th--84th percentile range shown as the shaded region in Figure~\ref{fig:lya_evolution}.
At $z\simeq6.2$, both of our measurements lie below the central value of this relation. 
The inner-region measurement overlaps with the broad uncertainty of the fitted evolution once its measurement uncertainty is taken into account, while the outer-region value lies systematically lower, although with a large uncertainty.

\section{ENVIRONMENTAL TRENDS IN GALAXY PROPERTIES}
\label{sec:galprops}

For the 63 HAEs satisfying the multi-band detection criteria described in Section~\ref{sec:data_sample},
we examine the stellar mass $M_*$, \Ha-based star-formation rate (SFR$_{\rm H\alpha}$), specific star-formation rate (sSFR), UV continuum slope $\beta$, dust reddening $E(B-V)$, stellar age, and $M_{\rm UV}$ as a function of $R_{\rm cen}$. 
For the 40 of these HAEs with reliable NB872 photometry, we also examine the individual \fesca\ and \ewLya\ derived in \citetalias{Shimizu+26}. 
We refer the reader to \citetalias{Shimizu+26} for details of the measurements of these quantities.
We quantify the monotonic relation between $R_{\rm cen}$ and each quantity using Kendall's rank correlation test. 
For \fesca\ and \ewLya, we adopt upper limits derived from the $3\sigma$ NB872 flux limits for NB872 non-detections 
and use the censored Kendall rank-correlation test following \citetalias{Shimizu+26}.

\begin{figure*}
    \centering
    \includegraphics[width=\textwidth]{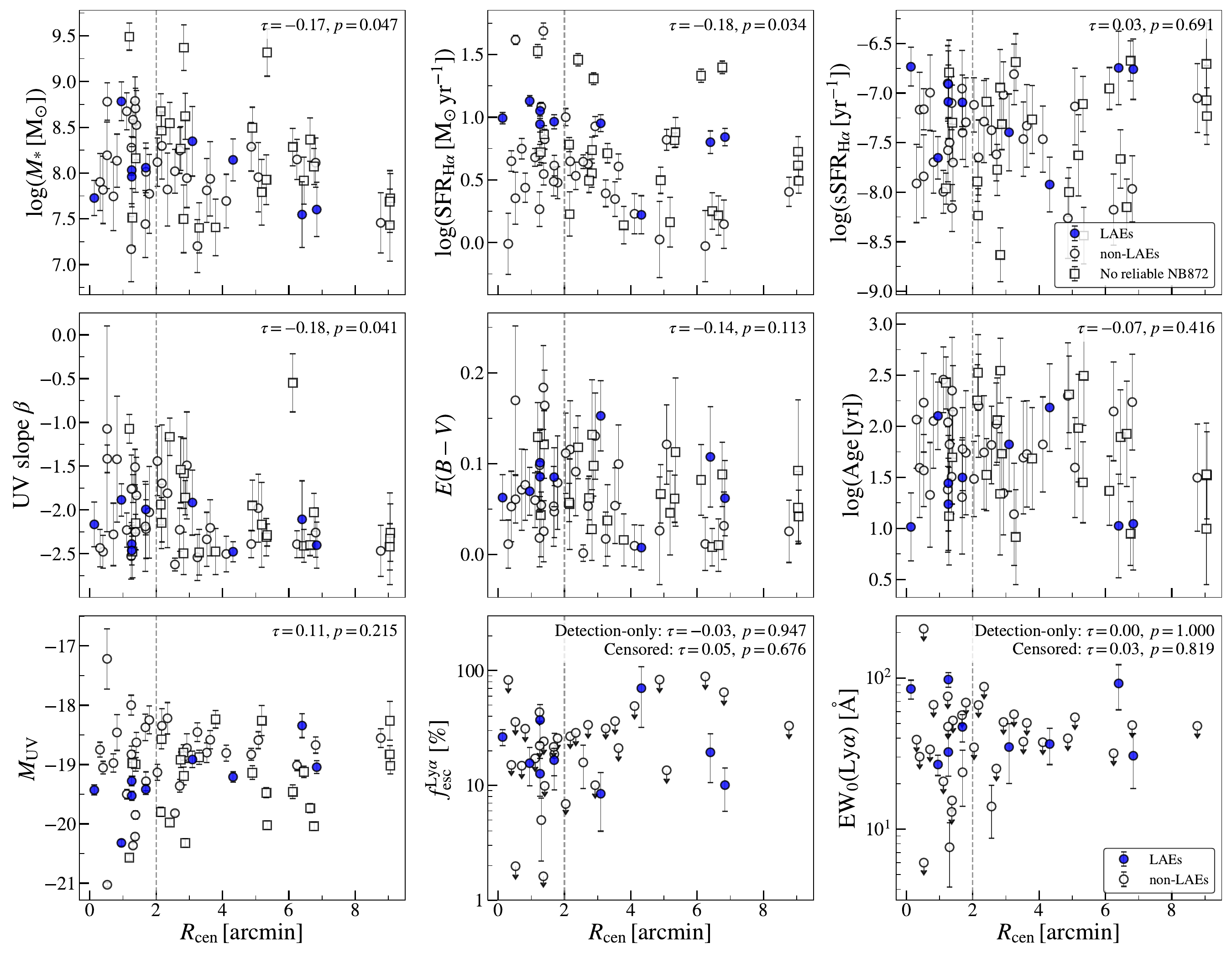}
    \caption{Galaxy properties as a function of $R_{\rm cen}$. From left to right and top to bottom, the panels show $M_*$, $\rm{SFR}_{H\alpha}$, sSFR, UV slope $\beta$, $E(B-V)$, stellar age, $M_{\rm UV}$, \fesca, and \ewLya. 
    The first seven panels show the 63 HAEs satisfying the multi-band
    detection criteria described in Section~\ref{sec:data_sample}.
    Filled blue circles indicate LAEs, open circles indicate non-LAEs,
    and open squares indicate sources without reliable NB872 photometry.
    The last two panels show the 40 HAEs with reliable NB872 photometry, using the same symbols for LAEs and non-LAEs.
    Downward arrows indicate the $3\sigma$ upper limits on \fesca\ and \ewLya\ for \Lya non-detections.
    The Kendall $\tau$ and $p$ values are shown in each panel, with both the detections-only and censored results given for \fesca\ and \ewLya.
    The vertical dotted line marks $R_{\rm cen}=2\arcmin$.
    }
    \label{fig:galaxy_properties}
\end{figure*}

Figure~\ref{fig:galaxy_properties} shows the galaxy properties as a function of $R_{\rm cen}$. 
We find weak negative correlations between $R_{\rm cen}$ and $M_\ast$ ($\tau=-0.17$, $p=0.047$), SFR$_{\rm H\alpha}$ ($\tau=-0.18$, $p=0.034$), and $\beta$ ($\tau=-0.18$, $p=0.041$), with all three $p$-values below 0.05.
Despite the substantial scatter visible in Figure~\ref{fig:galaxy_properties}, these trends indicate that galaxies closer to the density center thus tend to be more massive, have higher SFRs, and have redder UV slopes.
We confirm that these trends are qualitatively unchanged when
using alternative correlation statistics.
Although the environmental definitions and spatial scales differ from those adopted here, qualitatively similar trends have been reported at high redshift.
Using a fifth-nearest-neighbour density estimator, \citet{Li+25} found higher SFRs and redder UV slopes in denser environments.
At $z=7.88$, \citet{Witten+26} found that the core of the compact A2744-$z7p9$ protocluster preferentially hosts massive and dusty galaxies, while \citet{Osone+26} reported significant radial trends in stellar mass, SFR, and dust attenuation within the same structure.

The remaining galaxy properties, namely the sSFR, $E(B-V)$, stellar age, and $M_{\rm UV}$, show no significant dependence on $R_{\rm cen}$.
In particular, the sSFR does not increase toward the density center despite the higher SFRs there.
$E(B-V)$ tends to increase toward smaller $R_{\rm cen}$, similar to $\beta$, although the correlation is not significant.

Finally, neither the individual \fesca\ nor \ewLya\ shows a significant correlation with $R_{\rm cen}$ in either the censored or detection-only test.
Repeating the tests for the same $-20.25<M_{\rm UV}<-18.75$ sample adopted in Section~\ref{sec:lya_visibility} does not change this result.
Therefore, while the inner--outer comparison in Section~\ref{sec:lya_visibility} shows higher average \Lya visibility in the inner region, the strengths of \Lya emission for individual galaxies do not exhibit a significant dependence on $R_{\rm cen}$.

\section{Wide-field LAE Distribution around the HAE Overdensity}
\label{sec:wide_lae}

We examine the wide-field distribution of Ly$\alpha$-selected galaxies around the HAE overdensity. 
The HAE density center used in this analysis is determined solely from the spatial distribution of the 84 \Ha-selected galaxies described in Section~\ref{sec:densitymap}, independently of the LAE distribution.
We construct a wide-field LAE sample using the Subaru/HSC NB872 imaging centered on CEERS field, together with the HSC-$i$ and $z$ images retrieved from the latest data release of the Hyper Suprime-Cam Legacy Archive (HSCLA2020, \citealt{HSCLA}). 
Following the color selection criteria of \citet{Arita+26}, 
we adopt their criterion corresponding to \ewLya$\ge25$\,\AA\ to identify LAEs at $z\simeq6.2$, matching the LAE definition used throughout this work.
Regions strongly affected by bright stars and stray light
are masked and excluded from the subsequent spatial analysis.
All candidates are visually inspected to remove spurious detections and sources affected by imaging artifacts. 
This procedure yields 137 LAEs with NB872 magnitudes ranging from $23.7$ to $25.8$ over an effective HSC area of $1.4\,{\rm deg}^2$ after applying the masks described above.
This effective area is approximately 60 times larger than that of CEERS field.
We note that, because this LAE sample requires an NB872 detection at ${\rm S/N_{NB872}}\ge5$, some Ly$\alpha$-detected HAEs are not included in the sample.
Unlike F470N imaging, the depth varies significantly across the entire field in the NB872 image.
We therefore estimate the detection completeness for each $11'\times11'$ square region called a patch \citep{hscpipe}, 
by injecting artificial galaxies into the NB872 images and repeating the source-detection procedure, following \citet{Arita+26}.

\begin{figure*}
    \centering
    \includegraphics[width=\textwidth]{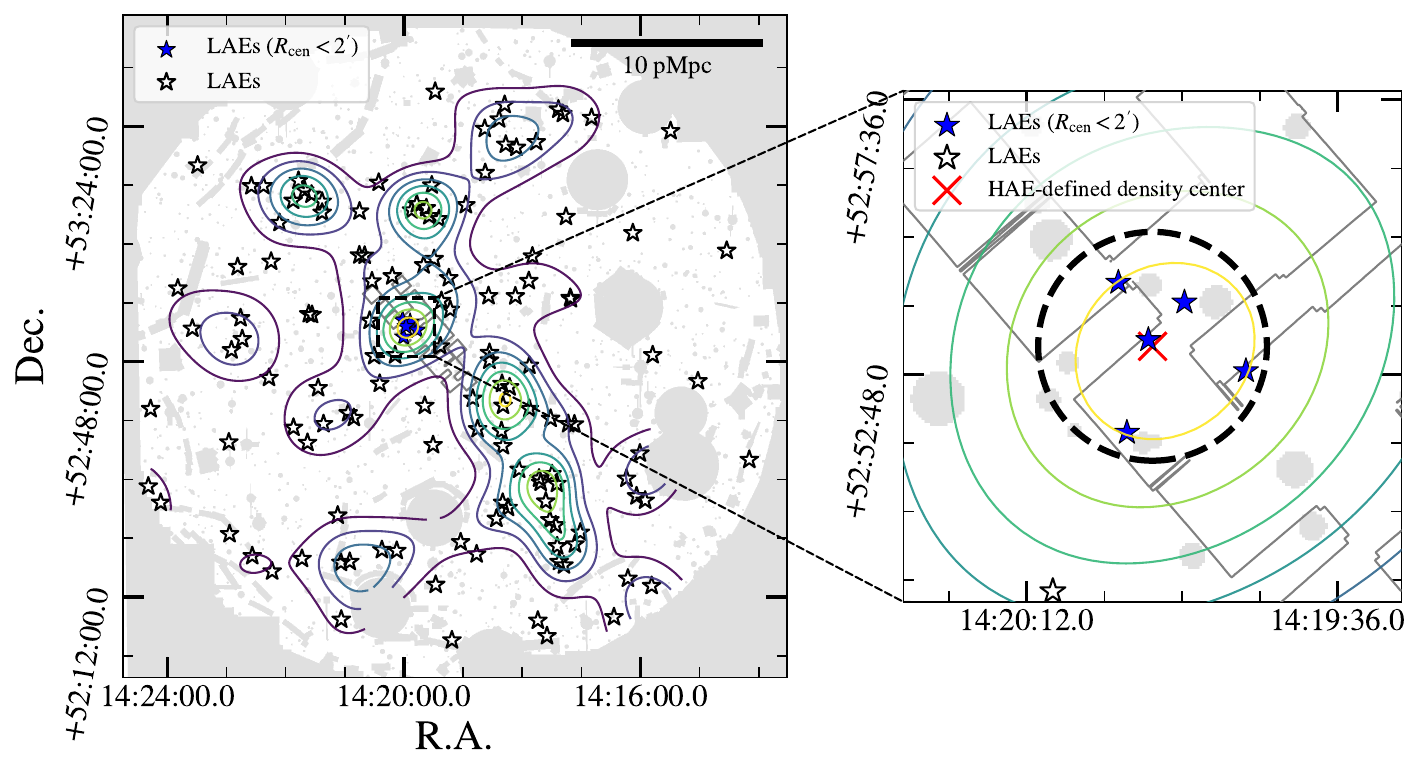}
    \caption{
    Wide-field distribution of the NB872-selected LAEs. 
    The left panel shows the full HSC field, while the right panel shows an enlarged view of the central region.
    In both panels, open stars show the LAEs, with filled blue stars highlighting those within $R_{\rm cen}<2\arcmin$ of the HAE-defined density center.
    The colored contours show the relative LAE overdensity, $\delta_{\rm LAE}$, at levels of $0.0$, $0.5$, $1.0$, $1.5$, $2.0$, $2.5$, and $3.0$.
    Gray regions indicate masked areas, while the gray outline in the center shows CEERS footprint.
    In the right panel, the HAE-defined density center and the $R_{\rm cen}=2\arcmin$ circle centered on it are shown.
}
    \label{fig:wide_lae}
\end{figure*}

Figure~\ref{fig:wide_lae} shows the spatial distribution of the selected LAEs over the wide HSC field. 
We construct the overdensity field using the same adaptive Gaussian KDE and edge-correction frameworks as for the HAE distribution in Section~\ref{sec:densitymap}, while weighting each LAE by the inverse of its local NB872 detection completeness to account for the spatial variation in imaging depth.
The pilot bandwidth is determined independently for the LAE sample, yielding $b_0=10.6$ cMpc ($=4\farcm4$ at $z=6.2$) for the Gaussian kernel.
We then define the relative LAE overdensity, $\delta_{\rm LAE}$, in the same manner as the HAE overdensity in Section~\ref{sec:densitymap}.

At the HAE-defined density center, the LAE overdensity reaches $\delta_{\rm LAE}=3.26$.
Moreover, the position of the maximum $\delta_{\rm LAE}$ across the entire $1.4\,{\rm deg}^{2}$ HSC field lies within $2'$ of the HAE-defined density center.
Thus, the prominent overdensity identified independently from the \Ha-selected population spatially coincides with the strongest overdensity traced by the wide-field LAE distribution.

\section{Discussion}
\label{sec:discussion}

The high \Lya visibility in an HAE overdense region
is particularly notable when compared with the environmental variation of the galaxy population. Both $X_{\rm LAE}$ and the stacked $f_{\rm esc}^{\rm Ly\alpha}$ are higher in the inner region than in the outer region with modest significance.
At the same time, galaxies closer to the density center tend to be more massive, have higher SFRs, and exhibit redder UV continuum slopes. 
Using the same parent HAE sample, \citetalias{Shimizu+26} found a significant dependence of \fesca\ on the UV continuum slope $\beta$, with lower \fesca\ toward redder $\beta$, consistent with \Lya photons escaping more efficiently along low-attenuation sightlines traced by bluer UV continua.
Since galaxies closer to the density center have systematically redder UV slopes, this trend would predict lower, rather than higher, \Lya escape toward the inner region.

In contrast, the higher SFR toward the density center could, in principle, favor higher \Lya escape in the inner region if it reflected systematically enhanced starburst activity, since stronger stellar feedback can facilitate the escape of \Lya photons by reducing the neutral-gas opacity \citep[e.g.][]{Kimm+19}.
However, the sSFR shows no significant dependence on $R_{\rm cen}$.
To further assess the star-formation activity relative to galaxies of similar stellar mass, we also examine $\Delta_{\rm SFMS}$, defined as the offset from the \Ha-based star-formation main sequence of \citet{Clarke+24}, and find no significant dependence on $R_{\rm cen}$ ($\tau=-0.046$, $p=0.598$).
This suggests that the SFR trend is largely associated with the increase in stellar mass. 
In addition, the $M_{\rm UV}$ distributions of the inner and outer samples show no statistically significant difference in a two-sample KS test,
indicating that the inner--outer difference in \Lya visibility is unlikely to arise simply from a systematic difference in UV luminosity.
Therefore, galaxy-property trends would not, by themselves, naturally produce higher \Lya visibility inside the overdensity.
In other words, 
while the underlying galaxy population becomes, on average, less favorable for \Lya escape toward the density center, both $X_{\rm LAE}$ and \fesca\ indicate the opposite trend.

The higher \Lya visibility in the HAE overdensity also contrasts with low-$z$ observations.
At $z=2.53$, \citet{Shimakawa+17} found a deficit of LAEs in the overdense regions traced by HAEs, together with a lower stacked \fesca\ in regions of higher HAE density.
\citet{Daikuhara+25} similarly reported that LAEs at $z=2.30$ preferentially avoid the overdense regions and filamentary structures traced by HAEs. 
These studies demonstrate that high densities of 
HAEs do not generally lead to enhanced \Lya visibility. 
Instead, neutral gas and dust in the ISM of galaxies residing in overdense environments can suppress the escape of \Lya photons.
The opposite trend observed in this study at $z\simeq6.2$ is therefore suggestive of an additional effect operating during the EoR.

A natural interpretation is that 
the transmission of \Lya photons through the neutral IGM is suppressed in regions outside the \Ha\ overdensity, while the \Lya attenuation may be reduced in overdense regions during the EoR.
Consistent with this picture, $X_{\rm LAE}$ and the stacked \fesca\ in the outer region 
are broadly consistent with other field measurements at $z\simeq6$, whereas the measurement in the inner region shows no comparable deficit, although the uncertainties are large, particularly for the outer region. 
Enhanced \Lya visibility in galaxy overdensities during the EoR has been attributed to weaker IGM attenuation within large ionized bubbles \citep[e.g.,][]{Endsley+22,Chen+26}.
The wide-field LAE distribution is also qualitatively consistent with this interpretation, with the HAE-defined density center coinciding with the strongest LAE overdensity across the wider HSC field.

The observational results of \citet{Zhu+26} appear to contrast with our findings: using \Ha\ and [O\,{\sc iii}] emitters identified with JWST/NIRCam wide-field slitless spectroscopy as tracers of the galaxy density field, they found lower \Lya visibility in overdense environments of line-emitters even at $z>6$.
Their radiative-transfer simulations show that the environmental dependence of \Lya IGM transmission can change sign as reionization progresses.
When the neutral fraction is still appreciable, \Lya emission from galaxies in overdense regions is less attenuated by the IGM, whereas this trend reverse as the IGM becomes more highly ionized.
Other suppressive effects in overdense environments may then become relatively more important.
Our results suggest that the HAE overdensity studied here is still at the stage in which \Lya emission from galaxies in overdense regions is less attenuated by the IGM.

\section{Conclusions}\label{sec:conclusion}

Using an \Ha-selected galaxy sample that defines the environment independently of \Lya emission, we find tentative evidence for higher \Lya visibility in the prominent HAE overdensity at $z\simeq6.2$ than in the surrounding region.
Both $X_{\rm LAE}$ and the stacked \fesca\ are higher in the inner region than in the outer region, although the statistical significance of these differences remains modest.
HAEs closer to the center are more massive and have redder UV slopes, trends that would not naturally explain the higher \Lya visibility.
These results favor a picture in which \Lya emission is less attenuated by the neutral IGM within the overdense region during the EoR.
Consistent with this picture, the HAE-defined density center coincides with the strongest LAE overdensity across the wider $1.4\,{\rm deg}^2$ HSC field.
This suggests that spatial variations in IGM attenuation remain important in shaping \Lya visibility near the end of reionization.

The present results remain limited by the small sample size. 
The inner--outer comparison of \Lya visibility is made based only on 21 HAEs.
In addition, the uncertain line-of-sight distribution may also bias the environmental comparison because 
NB872 probes a narrower redshift range than F470N. 
If the central overdensity lies near peak NB872 transmission while the outer HAEs span a broader redshift range, filter-wing attenuation could partly account for the higher \Lya visibility in the inner region.
Spectroscopic redshifts of the member galaxies will therefore be essential to establish the three-dimensional structure of the overdensity and to assess the impact of this effect.

More direct tests of the ionized-bubble interpretation will require probes of the IGM ionization state that are less sensitive to the
production and escape of \Lya photons within individual galaxies.
Ly$\alpha$-forest transmission, long used to constrain the ionization state of the high-redshift IGM (e.g., \citealt{Fan+06}) and now increasingly accessible along multiple galaxy sightlines with JWST/NIRSpec (e.g., \citealt{Meyer+25,Hu+26}), together with future 21-cm mapping of \Hi, will be important for constraining the presence, geometry, and extent of the ionized bubble.


\begin{acknowledgments}
We thank Yuma Sugahara, Beomchan Koh, Yuto Kuwayama for constructive discussions.
SS is supported by the Japan Society for the Promotion of Science (JSPS) KAKENHI grant number JP26KJ0916.
NK is supported by the Japan Society for the Promotion of Science through Grant-in-Aid for Scientific Research 25H00663, 25K01038, 25K01044.
JA is supported by the Japan Society for the Promotion of Science (JSPS) KAKENHI grant number JP24KJ0858 and International Graduate Program for Excellence in Earth-Space Science (IGPEES), a World-leading Innovative Graduate Study (WINGS) Program, the University of Tokyo.
AI is supported by JSPS KAKENHI Grant Number 23H00131, 26H02069, 25K00020, 24H00002.
KI acknowledges support from the Independent Research Fund Denmark (DFF) under grant 3120-00043B. 
The Cosmic Dawn Center (DAWN) is funded by the Danish National Research Foundation under grant No. 140.
SK is supported by the JSPS KAKENHI grant number 24KJ0058 and 24K17101.
MK is supported by the JSPS KAKENHI grant number JP25K01032.
KN acknowledges support from JSPS KAKENHI grant 20H00180, 24H00002, 24H00241, JP25K01032, and the JSPS International Leading Research (ILR) project, JP22K21349.
KN also acknowledges support from the Kavli IPMU, the World Premier Research Center Initiative (WPI), UTIAS, and the University of Tokyo.
MO is supported by the Japan Society for the Promotion of Science (JSPS) KAKENHI Grant Number 24K22894.
RS is also supported by the JSPS KAKENHI grant number JP25K01044. 

The Hyper Suprime-Cam (HSC) collaboration includes the astronomical communities of Japan and Taiwan, and Princeton University. The HSC instrumentation and software were developed by the National Astronomical Observatory of Japan (NAOJ), the Kavli Institute for the Physics and Mathematics of the Universe (Kavli IPMU), the University of Tokyo, the High Energy Accelerator Research Organization (KEK), the Academia Sinica Institute for Astronomy and Astrophysics in Taiwan (ASIAA), and Princeton University. Funding was contributed by the FIRST program from the Japanese Cabinet Office, the Ministry of Education, Culture, Sports, Science and Technology (MEXT), the Japan Society for the Promotion of Science (JSPS), Japan Science and Technology Agency (JST), the Toray Science Foundation, NAOJ, Kavli IPMU, KEK, ASIAA, and Princeton University. 

This paper makes use of software developed for Vera C. Rubin Observatory. We thank the Rubin Observatory for making their code available as free software at http://pipelines.lsst.io/.

This paper is based on data collected at the Subaru Telescope and retrieved from the HSC data archive system, which is operated by the Subaru Telescope and Astronomy Data Center (ADC) at NAOJ. Data analysis was in part carried out with the cooperation of Center for Computational Astrophysics (CfCA), NAOJ. 
This paper is also based in part on data from the Hyper Suprime-Cam Legacy Archive (HSCLA), which is operated by the Subaru Telescope. The original data in HSCLA was collected at the Subaru Telescope and retrieved from the HSC data archive system, which is operated by the Subaru Telescope and Astronomy Data Center at National Astronomical Observatory of Japan. 
We are honored and grateful for the opportunity of observing the Universe from Maunakea, which has the cultural, historical and natural significance in Hawaii. 

The Pan-STARRS1 Surveys (PS1) and the PS1 public science archive have been made possible through contributions by the Institute for Astronomy, the University of Hawaii, the Pan-STARRS Project Office, the Max Planck Society and its participating institutes, the Max Planck Institute for Astronomy, Heidelberg, and the Max Planck Institute for Extraterrestrial Physics, Garching, The Johns Hopkins University, Durham University, the University of Edinburgh, the Queen’s University Belfast, the Harvard-Smithsonian Center for Astrophysics, the Las Cumbres Observatory Global Telescope Network Incorporated, the National Central University of Taiwan, the Space Telescope Science Institute, the National Aeronautics and Space Administration under grant No. NNX08AR22G issued through the Planetary Science Division of the NASA Science Mission Directorate, the National Science Foundation grant No. AST-1238877, the University of Maryland, Eotvos Lorand University (ELTE), the Los Alamos National Laboratory, and the Gordon and Betty Moore Foundation.

This work is based in part on observations made with the NASA/ESA/CSA James Webb Space Telescope. The data were obtained from the Mikulski Archive for Space Telescopes (MAST) at the Space Telescope Science Institute, which is operated by the Association of Universities for Research in Astronomy, Inc., under NASA contract NAS 5-03127 for JWST. The JWST/NIRCam F090W and F470N observations from program \#2234 used in this work are available from MAST at \dataset[https://doi.org/10.17909/0pwe-j225]{https://doi.org/10.17909/0pwe-j225}, and the CEERS Public Data Release 1.0 products used in this work are available at \dataset[https://doi.org/10.17909/z7p0-8481]{https://doi.org/10.17909/z7p0-8481}.

We acknowledge the use of GitHub Copilot, ChatGPT (OpenAI), and Google Gemini for English-language editing, wording refinement, and translation of parts of the manuscript. 
The tools were used only to improve clarity and readability.
We reviewed all AI-assisted text and take full responsibility for the final manuscript.
\end{acknowledgments}

\facilities{HST(ACS), JWST (NIRCam), Subaru (HSC)}

\software{astropy \citep{2013A&A...558A..33A,2018AJ....156..123A,2022ApJ...935..167A}, matplotlib \citep{matplotlib}, numpy \citep{Numpy}
          }



\bibliography{ref}{}
\bibliographystyle{aasjournalv7}



\end{document}